\documentclass[pre,twocolumn,showpacs,superscriptaddress,preprintnumbers,floatfix]{revtex4}

\usepackage{dcolumn}
\usepackage{url}
\usepackage{amsmath}
\usepackage{amssymb}
\usepackage{graphicx}
\usepackage{bm}   
\usepackage{bbm}   
\usepackage{verbatim}
\usepackage{stmaryrd}
\usepackage{amsthm}

\def\bra#1{\mathinner{\langle{#1}|}}
\def\ket#1{\mathinner{|{#1}\rangle}}

\theoremstyle{break} 	
\theoremstyle{plain} 	
\theoremstyle{break} 	
\theoremstyle{plain} 	
\theoremstyle{break} 	
\theoremstyle{plain} 	
\theoremstyle{break} 	
\theoremstyle{plain}	
\theoremstyle{break}	 
\theoremstyle{plain}	

\newcommand{\Prob}      {\mathrm{Pr}}

\newcommand{\svector}[1]{\bra{ {#1} }}
\newcommand{\asympt}    {\rho^s }

\newcommand{\past}{\stackrel{\leftarrow}{S}}

\newcommand{\future}{\stackrel{\rightarrow}{S}}

\def\hmu   {h_\mu}

\def\Abet {\mathcal{A}}
\def\EE   {{\bf E} }

\def\TI   {{\bf T} }
\def\l2   {{\rm log}_2}

\begin{document}

\title{Intrinsic Quantum Computation}

\author{James P. Crutchfield}
\email{chaos@cse.ucdavis.edu}
\affiliation{Center for Computational Science \& Engineering and Physics Department,
University of California Davis, One Shields Avenue, Davis, CA 95616}

\author{Karoline Wiesner}
\email{karoline@cse.ucdavis.edu}
\affiliation{Center for Computational Science \& Engineering and Physics Department,
University of California Davis, One Shields Avenue, Davis, CA 95616}

\date{\today}

\bibliographystyle{unsrt}

\begin{abstract}
We introduce ways to measure information storage in quantum systems, using
a recently introduced computation-theoretic model that accounts for
measurement effects. The first, the quantum excess entropy, quantifies the
shared information between a quantum process's past and its future. The
second, the quantum transient information, determines the difficulty with
which an observer comes to know the internal state of a quantum process
through measurements. We contrast these with von Neumann entropy and quantum
entropy rate and provide a closed-form expression for the latter for the
class of deterministic quantum processes.
\end{abstract}

\preprint{Santa Fe Institute Working Paper 06-11-045}
\preprint{arxiv.org/quant-ph/0611202}

\maketitle




\vspace{-.25in}

Poincar\'e discovered that classical mechanical systems can appear to
be random \cite{Poin92}. Kolmogorov, adapting Shannon's theory of
communication \cite{Shan48a}, showed that their degree of randomness
can be measured as a rate of information production \cite{Kolm59}.
Shannon, in fact, adopted the word ``entropy'' to describe information
transmitted through a communication channel based on a suggestion by von
Neumann, who had recently used entropy to describe the distribution of
states in quantum systems \cite[Ch. 5]{Neum32a}. Information
has a long history in quantifying degrees of disorder
in both classical and quantum mechanical systems.

In a seemingly unrelated effort, Feynman proposed to develop quantum
computers \cite{feynman:82} with the goal of (greatly) accelerating
simulation of quantum systems. Their potential power, though, was brought
to the fore most recently by the discovery of algorithms that would run
markedly faster on quantum computers than on classical computers.
Experimental efforts to find a suitable physical substrate for a quantum
computer have been well underway for almost a decade now \cite{Zole05a}.

In parallel, the study of the quantum behavior of classically chaotic
systems gained much interest \cite{reic04a}, most recently including
the role of measurement. It turns out that measurement interaction
leads to genuinely chaotic behavior in quantum systems, even far from
the semi-classical limit \cite{habib:06}.

How are computing, information creation, and dynamics related? A contemporary
view of these three historical threads is that they are not so
disparate. We show here that a synthesis leads to methods to
analyze how quantum processes store and manipulate information---what
we refer to as \emph{intrinsic quantum computation}.

Computation-theoretic comparisons of classical (stochastic) and
measured quantum systems showed that a quantum system's
behavior depends sensitively on how it is measured. The differences
were summarized in a hierarchy of computational model classes for
quantum processes \cite{wiesner:06b}. Here we adopt an
information-theoretic approach that, on the one hand, is more quantitative
than and, on the other, is complementary to the structural view emphasized
by the computation-theoretic analysis. To start, recall the 
\emph{finite-state quantum generators} defined there. They consist of a finite
set of \emph{internal states} $Q = \{q_i: i = 0, \ldots, |Q|-1 \}$. The
\emph{state vector} over the internal states is an element of a
$|Q|$-dimensional Hilbert
space: $\bra{\psi} \in \mathcal{H}$. At each time step a
quantum generator outputs a symbol $s \in \Abet$ and updates its
state vector.

The temporal dynamics is governed by a set of $|Q|$-dimensional
\emph{transition matrices} $\{T(s) = U \cdot P(s), s \in \Abet \}$,
whose components are elements of the complex unit disk and where each
is a product of a unitary matrix $U$ and a projection operator $P(s)$.
$U$ governs the evolution of
the state vector. $\mathbf{P} =\{ P(s): s \in \Abet \}$ is a set of
\emph{projection operators} that determine how the state vector is measured.
We base our analysis on the class of projective measurements, applicable to
closed quantum systems\footnote{The generalization to open quantum systems
using any (including non-orthogonal) \emph{positive operator valued measures}
(POVM) requires a description 
of the state after measurement in addition to the measurement statistics.   
Repeated measurement, however, is the core of a quantum process as the term
is used here. We therefore leave the discussion of quantum processes observed
with a general POVM to a separate study.}. In the measurement setting used here the
output symbol $s$ is identified with the measurement outcome and
labels the system's eigenvalues. We represent the event of no measurement
with the symbol $\lambda$; $P(\lambda)$ can be thought of as the identity
matrix. Thus, starting with state vector $\svector{\psi_0}$, a single
time-step yields $\langle \psi(s) \vert = \langle\psi_0\vert U \cdot P(s)$.

To describe how an observer chooses to measure a quantum system we introduce
the notion of a \emph{measurement protocol}. A
\emph{measurement act}---applying
$P(m), m \in \Abet \bigcup \lambda$---returns a value $s \in \Abet$
or nothing ($m = \lambda$). A \emph{measurement protocol}, then, is a choice
of a sequence of measurement acts $m_t \in \Abet \bigcup \lambda$.
If an observer asks if the measurement outcomes $s_1 s_2 s_3$ occur,
the answer depends on the measurement protocol, since the observer
could choose protocol
$s_1 \lambda \lambda s_2 \lambda s_3$, $s_1 s_2 s_3$, or others.
Fixing a measurement protocol, then, the state vector after observing the
measurement series $m^N = m_1 \ldots m_N$ is
$
\langle \psi(m^N) \vert   
  = \langle\psi_0\vert 
  U \cdot P(m_1) \cdot U \cdot P(m_2) \cdots U \cdot P(m_N)
$.

A {\em quantum process} is the joint distribution
${\rm Pr} ( \ldots S_{-2} S_{-1} S_0 S_1 \ldots)$ over the infinite chain of
measurement random variables $S_t$. Defined in this way, it is the quantum
analog of what Shannon
referred to as an information source \cite{cover}. Starting a generator
in $\langle\psi_0\vert$ the probability of output $s$ is given by
the state vector without renormalization:
$\Prob(s) =  \left\Vert \psi(s) \right\Vert^2$. By extension,
the \emph{word distribution}, the probability of
outcomes $s^L$ from a sequence of $L$ measurements, is
$\Prob(s^L) =  \left\Vert \psi(s^L) \right\Vert^2$.

We can use the observed behavior, as reflected in the word distribution,
to come to a number of conclusions about how a quantum process generates
randomness and stores and transforms historical information. The
{\em Shannon entropy} of length-$L$ sequences is defined
\begin{align}
\label{eq:HL}
H(L)  &\equiv  - \sum_{ s^L \in {\cal A}^L } \Prob (s^L) \l2 \Prob (s^L) ~.
\end{align}
It measures the average surprise in observing the ``event'' $s^L$.
Ref. \cite{crutchfield:03} showed that a stochastic process's informational
properties can be derived systematically by taking derivatives and then
integrals of $H(L)$, as a function of $L$.
For example, the {\em source entropy rate} $\hmu$ is the rate of increase with
respect to $L$ of the Shannon entropy in the large-$L$ limit:
\begin{equation}
    \hmu \equiv \lim_{L \rightarrow \infty} \left[ H(L) - H(L-1) \right] \; ,
\label{ent.def}
\end{equation}
where the units are \emph{bits/measurement} \cite{cover}. The entropy rate
$h_\mu$ quantifies the irreducible randomness in measurement sequences
produced by a process: the randomness that remains after the correlations and
structures in longer and longer sequences are taken into account.
The latter, in turn, are measured by two complementary quantities.
The amount $I(\past;\future)$ of mutual information \cite{cover} shared
between a process's past $\past$ and its future $\future$ is given by the
\emph{excess entropy} $\EE$ \cite{crutchfield:03}. It is the subextensive
part of $H(L)$: 
\begin{equation}
  \EE = \lim_{L \rightarrow \infty} [ H(L) - \hmu L ]\;.
\label{EEfromEntropyGrowth}
\end{equation}
Note that the units here are \emph{bits}.
Ref. \cite{crutchfield:03} also showed that the amount of information an
observer must extract from measurements in order to know the internal state
is given by the \emph{transient information}: 
\begin{equation}
\TI \equiv \sum_{L=0}^\infty \left[ \EE + \hmu L - H(L) \right] \;,
\label{T.def}
\end{equation}
where the units are {\em bits $\times$ measurements}.

If one can determine the word distribution $\Prob (s^L)$, then, in principle
at least, one can calculate a quantum process's informational properties:
$h_\mu$, $\EE$, and $\TI$. Fortunately, there are several classes of quantum
process for which one can give closed-form expressions. For example, we will
provide a way of computing $h_{\mu}$ exactly, using the finite-state machine
representation of a quantum process, similar to what is known for classical
stochastic processes \cite{Crut97a}. We then give examples of various
quantum processes at the end, measuring their intrinsic computation.

In quantum theory one distinguishes between complete and incomplete
measurements. A \emph{complete measurement} projects onto a one-dimensional
subspace of $\mathcal{H}$. A \emph{complete quantum generator} (CQG) is
simply a quantum generator observed with complete measurements.

Another, as it turns out, more general class of quantum processes are those
that can be described by a \emph{deterministic quantum generator} (DQG),
where each matrix $T(s)$ has at most one nonzero entry per row.
The importance of \emph{determinism} comes from the fact that it guarantees
that an internal-state sequence $q_t q_{t+1} q_{t+2} \ldots$ is in 1-to-1
correspondence with a measurement sequence $s_t s_{t+1} s_{t+2}$.
\footnote{Determinism here refers, as it does in automata theory\cite{hopcroft},
to the stated property; it does \emph{not} imply ``non-stochastic''.}

It simplifies matters if the word distribution is independent of the initial
state vector $\bra{\psi_0}$. This is achieved by switching to the density matrix
formalism \cite{wiesner:06b}. A \emph{stationary state distribution} $\rho^s$
can then be found for DQGs and
$\Prob(s^L) = \mathrm{Tr} \left[ T^{\dagger}(s^L)\rho^s T(s^L) \right]$
is start-state independent.

We showed that every DQG has an equivalent deterministic classical generator that produces
the same stochastic process \cite{wiesner:06b}. Specifically,
given a DQG $M = \{U,P(s)\}$, the \emph{equivalent} classical generator
$M^{\prime} = \{T,P(s)\}$ has unistochastic transition matrix
$T_{ij} = | U_{ij} |^2$.

As a consequence, when a quantum process can be represented by a complete
or a deterministic generator, closed-form expressions exist for several of
the information quantities. For example, adapting Ref. \cite{Crut97a} to
DQGs we obtain the \emph{quantum entropy rate}:
\begin{equation}
h_\mu = - |Q|^{-1} \sum_{i=0}^{|Q|-1}\sum_{j=0}^{|Q|-1}
	|U_{ij}|^2 \log_2 |U_{ij}|^2 ~,
\end{equation}
using the fact that $\rho^s_{ii}=|Q|^{-1}$ for DQGs. This should be compared
to the entropy rates for nondeterministic classical and quantum processes
which, in general, have no closed-form expression.
For DQGs we introduce the
\emph{internal-state quantum entropy}:
\begin{equation}
S_q = - \sum_{i=0}^{|Q|-1} \rho^s_{ii} \log_2 \rho^s_{ii} ~,
\end{equation}
which measures the average uncertainty in knowing the internal state. Again,
$\rho_{ii}=|Q|^{-1}$ for DQGs allows us to simplify: $S_q = \log_2 |Q|$.

$S_q$, as defined here, is nothing other than
the \emph{von Neumann entropy} $S(\rho)$ of the density matrix $\rho$
\cite{nielsen}. 
Note, that our use of the density matrix is as a time average, not an ensemble
average. This should be further compared to the 
von Neumann entropy
$S(\otimes_{i=0}^{L-1} \rho_i)$ over a sequence of density matrices produced
by a stationary quantum source \cite{wehrl:78}. Using this, a
\emph{density-matrix quantum entropy rate} can be defined as the limit of
$S(\otimes_{i=0}^{L-1} \rho_i)/L$ for large $L$. Note that these alternative
definitions of entropy and entropy rate suffer from two problems. First,
they refer to internal variables that are not directly measurable.
Second, they do not take the effects of measurement into account.
Importantly, $\hmu$, $\EE$, and $\TI$ do not suffer from these problems.
Let's explore their consequences for characterizing intrinsic computation
in several simple quantum dynamical systems. (An exhaustive analysis of
all (deterministic and non-deterministic) few-qubit quantum processes will
appear elsewhere.)

\begin{figure}  
\begin{center}
\resizebox{!}{1.50in}{\includegraphics{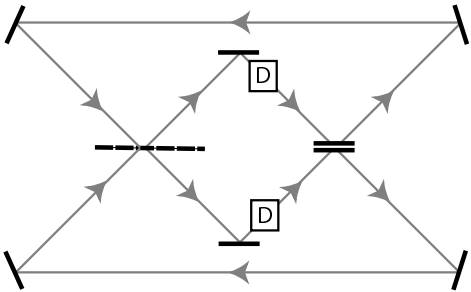}}
\end{center}
\caption{Iterated beam splitter: Thick solid lines are mirrors. Photon
  detectors, marked as D, are placed in upper and lower photon paths
  (solid gray lines).
  }
\label{fig:IteratedBeamSplitter}
\end{figure}

The \emph{iterated beam-splitter} (Fig.~\ref{fig:IteratedBeamSplitter})
is a quantum system that, despite its simplicity, makes a direct connection
to familiar experiments. Photons are sent through a beam-splitter,
producing two possible paths, which are redirected by mirrors and recombined
at the beam-splitter after passing around the feedback loop determined by the
mirrors. Nondestructive detectors are located along the upper and lower paths.

The iterated beam splitter is a quantum dynamical system with
a two-dimensional state space
(which path is taken around the loop) with eigenstates ``above''
$\svector{\phi_A}$ and ``below''$\svector{\phi_B}$. Its dynamics
are given by a unitary operator for the beam-splitter, the Hadamard matrix
$
U_H = \tfrac{1}{\sqrt{2}}
    \left(
	\begin{smallmatrix}
	1 & 1 \\
	1 & -1
	\end{smallmatrix}
	\right) $,
and measurement operators representing the detectors:
$P(0) = \ket{0}\bra{0}$ and $P(1) = \ket{1}\bra{1}$,
in the experiment's eigenbasis. Measurement symbol $0$
stands for ``above'' and symbol $1$ stands for ``below''.

The detectors after
the first beam splitter detect the photon in the upper or lower path
with equal probability. Once the photon is measured, though, it is in
that detector's path with probability $1$. And so it enters the beam
splitter again in the same state as when it first entered. Thus,
the measurement outcome after the second beam splitter will have
the same uncertainty as after the first: the detectors still report
``above'' or ``below'' with equal probability. The resulting sequence
of detector outcomes after many circuits of the feedback loop is simply
a random sequence. Call this \emph{measurement protocol I}.

Now alter the experiment slightly by activating the detectors only after
every other circuit of the feedback loop. In this set-up, call it
\emph{protocol II}, the photon enters the first beam splitter, passes an
inactive detector and interferes with itself when it returns to the beam
splitter. This,
as we will confirm, leads to destructive interference of one path after
the beam splitter. The photon is thus in the same path after the second
visit to the beam splitter as it was on the first. The now-active detector
therefore reports with probability $1$ that the photon is in the upper path,
if the photon was initially in the upper path. If it was initially in the
lower path, then the detector reports that it is in the upper path with
probability $0$. The resulting sequence of path detections is a very
predictable sequence, compared to the random sequence from protocol I.
Note that both protocols are complete measurements.

We now construct a complete quantum generator for the iterated-beam splitter.
The output alphabet consists of two symbols denoting detection ``above'' or
``below'': $\Abet = \{0,1\}$. There are two internal states ``above'' and
``below'', each associated with one of the two system eigenstates:
$Q = \{A, B\}$. The transition matrices are $T(0) = U_H  P(0)$ and
$T(1) = U_H  P(1)$. We assume that the quantum process has been operating
for some time and so take
$\asympt = |Q|^{-1} \sum_{i\in Q}\ket{\phi_i}\bra{\phi_i}$.

One can readily verify that this representation of the iterated beam
splitter is a DQG. And so we can determine its classical equivalent
generator has transition matrices
$
T(0) = \frac{1}{2}
	\left(
	\begin{smallmatrix}
	1 & 0 \\
	1 & 0
	\end{smallmatrix}
	\right)
  ~\mathrm{and}~
T(1) = \frac{1}{2}
	\left(
	\begin{smallmatrix}
	0 & 1 \\
	0 & 1
	\end{smallmatrix}
	\right) $.
The sequences it generates for protocol I are described by the
uniform distribution at all lengths: $\Prob(s^L) = 2^{-L}$.

Note, however, that the probability distribution of the sequences for the
classical generator under protocol II is still the uniform distribution
for all lengths $L$. This could not be more different from the
behavior of the (quantum) iterated beam splitter under protocol II.
The classical generator is simply unable to capture the interference
effects present in this case. A second classical generator must be
constructed from the quantum generator's transition matrices for protocol
II. One finds
$
T(0) = \tfrac{1}{2}
	\left(
	\begin{smallmatrix}
		1 & 0 \\
		0 & 0
	\end{smallmatrix}
	\right)
	~\mathrm{and}~
T(1) = \tfrac{1}{2}
	\left(
	\begin{smallmatrix}
	0 & 0 \\
	0 & 1
	\end{smallmatrix}
	\right) $.
Starting the photon in the upper path again, for protocol II one finds
$\Prob(00\ldots) = \Prob(11\ldots) = 1/2$ and all other words have
zero probability.

Table \ref{tab:info} gives the informational quantities for the iterated
beam splitter under the protocols. Under protocol I it is maximally random
($\hmu = 1$), as expected. It also, according to $\EE = 0$, does not
store any historical information and the observer comes to know this
immediately ($\TI = 0$). In stark contrast, under protocol II, the iterated
beam splitter is quite predictable ($\hmu = 0$) and stores one bit of
information ($\EE = 1$)---whether the measurement sequence is $000\ldots$
or $111\ldots$. Learning which requires extracting one bit ($\TI = 1$)
of information from the measurements.

Note that the internal-state (von Neumann) entropy $S_q = 1$ under both
protocols: there are two equally likely states in both cases. It simply
reflects the single qubit in the iterated beam splitter, not whether that
qubit is useful in supporting intrinsic computation.

\begin{table}[tbp]
\begin{tabular}{|c||c|c||c|c|}
\hline
 Quantum & \multicolumn{2}{c||}{Iterated Beam} & \multicolumn{2}{c|}{Spin-1} \\
Dynamical System        & \multicolumn{2}{c||}{Splitter}      & \multicolumn{2}{c|}{Particle } \\
\hline
Protocol & ~~I~~         & II                     & I             & II \\
\hline
\hline
 $h_\mu$ [\emph{bits/measurement}] & 1         & 0          & 0.666	& 0.666 \\
 $S_q$ [\emph{bits}] & 1         & 1                      & 1.585 & 1.585 \\
 $\EE$   [\emph{bits}] & 0         & 1                      & 0.252 & 0.902 \\
 $\TI$   [\emph{bits$\times$measurement}] & 0         & 1   & 0.252 & 3.03 \\
\hline
\end{tabular}
\caption{Information storage and generation for example quantum processes:
  entropy rate $h_\mu$, internal-state entropy $S_q$, excess entropy $\EE$,
  and transient information $\TI$.
  }
\label{tab:info}
\end{table}

Now consider a second, more complex example: A spin-$1$ particle subject to
a magnetic field that rotates its spin. The state evolution can be described
by the unitary matrix
$
U = \left(
\begin{smallmatrix}
		1/\sqrt{2} & 1/\sqrt{2} & 0
		\\ 0 & 0 & -1
		\\ -1/\sqrt{2} & 1/\sqrt{2} & 0 
\end{smallmatrix}
	\right)
$,
which is a rotation in $\mathbb{R}^3$ around the y-axis by angle
$\tfrac{\pi}{4}$ followed by a rotation around the x-axis by $\tfrac{\pi}{2}$.

Using a suitable representation of the spin operators $J_i$ \cite[p.199]{peres},
such as:
$
J_x = \left(
	\begin{smallmatrix}
        0 & 0 & 0 \\
        0 & 0 & i \\
        0 & -i & 0
	\end{smallmatrix}
    \right)
$,
$
J_y = \left(
	\begin{smallmatrix}
        0 & 0 & i \\
        0 & 0 & 0 \\
        -i & 0 & 0
	\end{smallmatrix}
    \right)
$, and
$
J_z = \left(
	\begin{smallmatrix}
        0 & i & 0 \\
        -i & 0 & 0 \\
        0 & 0 & 0
	\end{smallmatrix}
    \right) ,
$
the relation $P_i = 1 - J_i^2$
defines a one-to-one correspondence between the projector $P_i$ and the
square of the spin component along the $i$-axis. The resulting measurement
poses the yes-no question, Is the square of the spin component along
the $i$-axis zero?

Define two protocols, this time differing in the projection operators. First,
consider measuring $J_y^2$; call this \emph{protocol I}. Then $U$ and the
projection operators
$P(0) = \ket{100}\bra{100} + \ket{001}\bra{001}$ and $P(1) = \ket{010}\bra{010}$
define a quantum generator.

The stochastic language produced by this process is the so-called
\emph{Golden-Mean Process} language \cite{crutchfield:03}. It is
defined by the set of \emph{irreducible forbidden words}
$\mathcal{F} = \{00\}$. That is, all measurement sequences occur,
except for those with consecutive $0$s. For
the spin-$1$ particle this means that the spin component along the
y-axis never equals 0 twice in a row. We call this
\emph{short-range correlation} since there is a correlation between a
measurement outcome at time $t$ and the immediately preceding one
at time $t-1$. If the outcome is $0$, the next outcome will be
$1$ with certainty. If the outcome is $1$, the next measurement is
maximally uncertain: outcomes $0$ and $1$ occur with equal probability.

Second, consider measuring the observable $J_x^2$; call this
\emph{protocol II}. Then $U$ and $P(0) = \ket{100}\bra{100}$ and
$P(1) = \ket{010}\bra{010} + \ket{001}\bra{001}$ define a quantum finite-state
generator. The stochastic language generated is the \emph{Even Process}
language \cite{crutchfield:03}. It is characterized by an infinite set
of irreducible forbidden words $\mathcal{F} = \{01^{2k-1}0\}, k=1,2,3,...$.
That is, if the spin component equals $0$ along the $x$-axis, it will be zero
an \emph{arbitrary large}, even number of consecutive measurements before
being observed to be nonzero. 

Table \ref{tab:info} gives the intrinsic-computation analysis
of the spin-$1$ system. Comparing it to the iterated beam
splitter, it's clear that the spin-$1$ system is richer---a
process that does not neatly fall into one or the other extreme of
exactly predictable and utterly unpredictable. In fact, under
protocols I and II it appears equally unpredictable: $\hmu \approx
0.333$. As before the von Neumann entropy is also the same
under both protocols. The amount of information that the spin-$1$
system communicates from the past to the future is nonzero,
with the amount under protocol I ($\EE \approx 0.252$) being
less than under protocol II ($\EE \approx 0.902$). These accord
with our observation that in the latter case there is a kind
of infinite-range temporal correlation. Not too much information
($\TI \approx 0.252$) must be extracted by the observer under protocol
I in order to see the relatively little memory stored in this
process. Interestingly, however, under protocol II this is
markedly larger ($\TI \approx 3.03$), indicating again that
the observer must extract more information to see just how
this process is monitoring ``evenness''.

We close by looking at a repeatedly measured quantum system in the context
of recent developments in quantum computation. Quantum control theory---a
paradigm of repeated classical-state measurement and feedback control---has
only recently been implemented as a means to drive quantum systems
toward a desired state or dynamic \cite{geremia:04}. Current
implementations are finite-dimensional and aim at the control of a known
quantum state. Therefore, our formalism can be used to characterize the
intrinsic computation of these quantum systems. In fact, it can be used on
any quantum system observed over time, whether its exact state is known or
not. If it is known and the applied measurement protocol generates a 
deterministic quantum process the above measures of intrinsic computation
can be calculated exactly using its quantum generator representation. Generally,
though, the intrinsic quantum computation of an unknown quantum state can be
measured experimentally by recording a time series of measurement outcomes
and computing the probability distribution. In fact, from the probability
distribution of a discrete time series of measurements one can compute the
intrinsic computation of both open and closed quantum systems.

We introduced several new information-theoretic quantities that
reflect a quantum process's intrinsic computation: its information
production rate, how much memory it apparently stores in generating
information, and how hard it is for an observer to synchronize.
We discussed how several of these informational quantities are related
to existing notions of entropy in quantum theory. The contrast
allowed us to demonstrate how much more they tell one about the
intrinsic computation supported by quantum processes and to
highlight the crucial role of measurement. For example, simply
knowing that a quantum process is built out of some number of qubits
only gives an upper bound on the possible information processing.
It does not reflect how the quantum system actually uses the qubits
to process information nor how much of its information
processing can be observed. For these, one needs the quantum
$\hmu$, $\EE$, and $\TI$.


UCD and the Santa Fe Institute supported this work via the Network
Dynamics Program funded by Intel Corporation. The Wenner-Gren Foundations,
Stockholm, Sweden, provided KW's postdoctoral fellowship.


\vspace{-.25in}

\bibliography{ref}

\end{document}